\documentclass{article}
\usepackage[english]{babel}
\usepackage{csquotes}
\usepackage{authblk}
\date{} % This will omit the date
\providecommand{\keywords}[1]{\textbf{Keywords:} #1}
\usepackage[top=1in, bottom=1in, left=1in, right=1in]{geometry} % USE 1 inch margins
\usepackage{multirow}

\usepackage{amsmath,amssymb,amsfonts} % for \mathbb
\usepackage{dsfont}                   % for \mathds{1} (blackboard 1)
\usepackage{graphicx}

\usepackage{subcaption}

\usepackage[hidelinks]{hyperref}
\usepackage{abstract}
\usepackage{setspace}
\usepackage[
  backend=biber,
  style=authoryear,
  sorting=nyt,
  maxcitenames=2,   % 1 author → Last; 2 authors → Last and Last
  mincitenames=1,   % 3+ authors → Firstauthor et al.
  maxbibnames=99,   % full author list in the References
  giveninits=true,  % initials in the References (not citations)
  uniquename=false, % don't inject given names in citations
  uniquelist=false, % don't expand author lists for disambiguation
  useprefix=true    % keep particles (e.g., "Van der") with surname
]{biblatex}

\usepackage[table]{xcolor}
\usepackage{booktabs}

\usepackage{enumitem}

\usepackage{doi}

\title{StrengthLawExtractor: A Fiji plugin for 3D morphological feature extraction from X-ray micro-CT data}

\author[1]{Qinyi Tian}
\author[1,*]{Laura E. Dalton} 

\affil[1]{Duke University, Department of Civil \& Environmental Engineering, Durham, NC, USA}

\affil[*]{Corresponding author email: laura.dalton@duke.edu}

\begin{document}

% \linenumbers 
% \switchlinenumbers

\maketitle
\doublespacing
\begin{abstract}

Non-destructive methods are essential for linking the microstructural geometry of porous materials to their mechanical behavior, as destructive testing is often infeasible due to limited material availability or irreproducible conditions. Micro-computed tomography (micro-CT) provides high resolution three dimensional reconstructions of porous microstructures, enabling direct quantification of geometric descriptors. Recent advances in morphometric theory have demonstrated that four independent morphometric measures (porosity, surface area, mean curvature, and Euler characteristic) are required to capture the relationship between microstructure and strength, thereby forming the basis of generalized strength laws. To facilitate practical application of this framework, a Fiji plugin was developed to extract the four morphometric measures (porosity, surface area, mean curvature, Euler characteristic) from micro-CT datasets automatically. The plugin integrates within the Fiji platform to provide reproducible, accessible, and user friendly analysis. The application of the tool demonstrates that the extracted descriptors can be readily incorporated into constitutive models and machine learning workflows, enabling the forward prediction of stress-strain behavior as well as the inverse design of microstructures. This approach supports non-destructive evaluation, accelerates materials selection, and advances the integration of imaging with predictive modeling in porous media research.

\vspace{5mm}

\end{abstract}

\vspace{5mm}

\newpage
\doublespacing
\section{Introduction}
\begin{flushleft}

%% Materials Characterization: https://www.sciencedirect.com/journal/materials-characterization

Non-destructive imaging techniques have become indispensable for the characterization of materials, particularly when linking microstructural features to macroscopic mechanical behavior \parencite{doi:10.1179/095066003225010254, doi:10.1177/1687814020913761}. Conventional destructive testing often suffers from limitations such as scarcity of suitable or representative specimens, the inability to replicate natural or service loading conditions, and the permanent loss of samples once they have been tested \parencite{DWIVEDI20183690}. For geological materials, accessing and extracting specimens from large depths is prohibitively expensive and remains impossible at certain depths. X-ray micro–computed tomography (CT) enables three dimensional reconstructions of pore space and solid networks without damaging the specimen \parencite{KETCHAM2001381, CNUDDE20131, WILDENSCHILD2013217, Dalton2020}. In biological materials, in vivo analysis requires techniques that preserve the integrity of tissues while X-ray micro–CT is widely used for non-destructive assessment of bone and small animal models \parencite{https://doi.org/10.1002/jbmr.141, Beaucage2016}. These limitations highlight the need for methods capable of extracting microstructural information without damaging the material.

X-ray micro-CT has emerged as one of the most effective tools for non-destructive three dimensional imaging of porous media \parencite{Wevers2018}. The ability to reconstruct volumetric microstructures with high spatial resolution provides direct access to critical descriptors such as porosity, connectivity, and grain morphology \parencite{Torquato2002}. This capability has transformed the study of porous media, ranging from rocks and soils to ceramics, composites, and biological tissues, enabling systematic investigations of the relationships between microstructural features and mechanical response \parencite{BLUNT2013197, Maire01012014}.  

Porous materials exhibit multiphysics and multiscale behavior, yet significant progress has been made in modeling their macroscopic strength using a reduced number of structural descriptors \parencite{Horstemeyer2010, Fish2021}. Early theoretical developments, such as the poromechanics framework and the Hall-Petch relation of Biot, emphasized porosity and grain size as key parameters governing mechanical resistance \parencite{Guevel2021}. Experimental studies further demonstrated exponential relationships between strength and porosity, as well as scaling effects linked to individual grain dimensions \parencite{RICE1989215, HAMID2021101833}. Building upon these foundations, morphometric theory established four independent geometric measures (i.e., geometric morphometers) are sufficient to capture the full influence of microstructural geometry on strength. 
% Based on Hadwiger’s theorem, where the strength of porous materials was linked to microstructural attributes through the strength law shown the general form of this dependence can be expressed as the strength law 
Invoking Hadwiger’s theorem for motion invariant, continuous, and additive valuations, the dependence of strength on microstructure admits a general representation, referred to here as the strength law \parencite{GUEVEL2022111454},
\begin{equation}
\sigma_{s} = \sigma_{s}^{*} f\left(\frac{M_{0}}{M_{0}^{*}}, \frac{M_{1}}{M_{1}^{*}}, \frac{M_{2}}{M_{2}^{*}}, \frac{M_{3}}{M_{3}^{*}}\right),
\end{equation}
where $\sigma_{s}$ is the macroscopic strength, $\sigma_{s}^{*}$ a reference strength, $M_{i}$ the morphometric measures, and $M_{i}^{*}$ the corresponding reference values determined by material type and environmental conditions.  

Recent advances in machine learning have expanded the opportunities to leverage the robust digital resources X-ray micro-CT data provides. Forward models allow prediction of stress-strain responses from microstructural descriptors \parencite{Lindqwister2025}, while inverse models provide pathways to design microstructures tailored to desired mechanical properties \parencite{Tian2025}. Both approaches require accurate, efficient, and reproducible quantification of the relevant morphometric features. This highlights the need for the development of tools to bridge imaging data with predictive models by extracting descriptors directly from reconstructed volumes in a robust and standardized manner.  

To meet this requirement, a Fiji plugin, \texttt{StrengthLawExtractor}, has been developed to compute the four essential morphometric measures from micro-CT datasets. Fiji \parencite{Schindelin2012Fiji} is a widely used, open source platform in both scientific and engineering communities, making it an ideal environment for deploying such a tool. The plugin provides automated and user friendly extraction of morphometric descriptors, ensuring reproducibility across studies and accessibility for a broad user base. By enabling direct quantification of microstructural parameters, the plugin supports the calibration of morphometric strength laws, facilitates the integration of imaging with machine learning workflows, and ultimately promotes non-destructive pathways for material characterization, selection, and design.

\end{flushleft}

\section{Quick Start Overview for Users}
\begin{figure}[htbp]
  \centering
  \begin{subfigure}{0.5\linewidth}
    \centering
    \includegraphics[width=\linewidth]{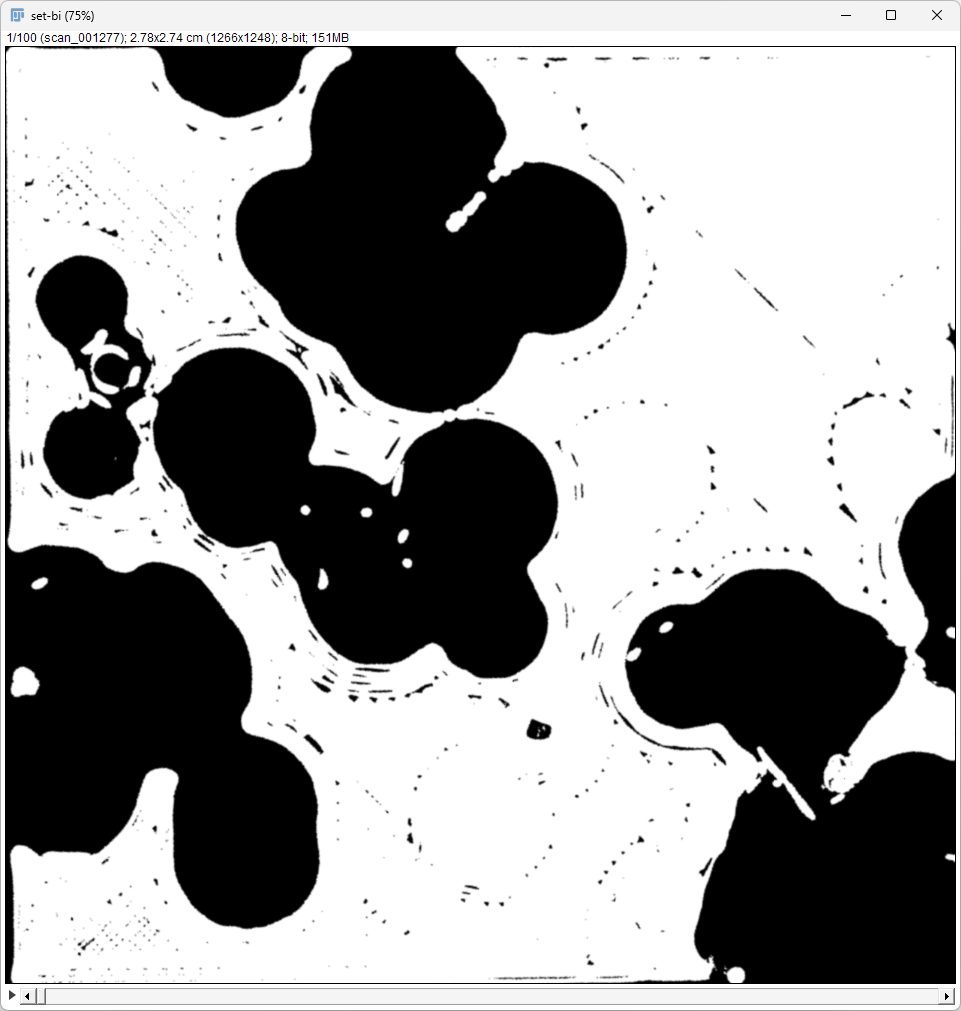}
    \caption{ }
    \label{fig:img1}
  \end{subfigure}\hfill
  \begin{subfigure}{0.497\linewidth}
    \centering
    \includegraphics[width=\linewidth]{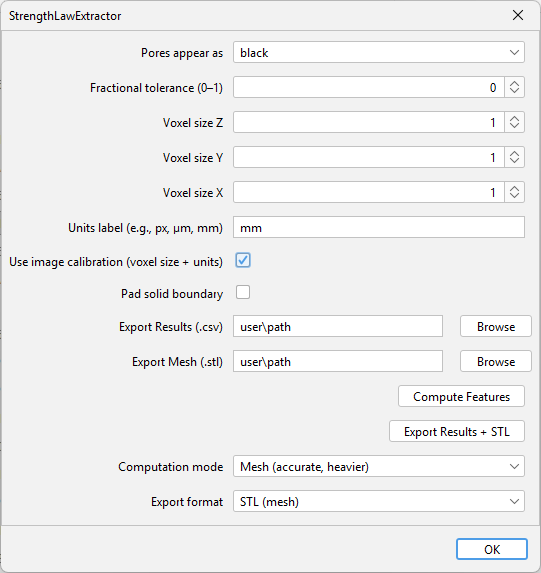}
    \caption{ }
    \label{fig:img2}
  \end{subfigure}
  \caption{(a) Example cross-sectional view of binarized input data; (b) User interface.}
  \label{fig:imgs1}
\end{figure}

\begin{flushleft}

\noindent \textbf{Plugin Function.} \texttt{StrengthLawExtractor} requires a 3D image stack to be imported, binarized, and then reports four 3D descriptors (i.e., the four geometric morphometers needed for the strength law shown in Equation 1) of the pore phase - porosity, surface area, mean curvature, and Euler characteristic - together with optional exports of the interface geometry.

\noindent \textbf{Mode Selection.}
Use \emph{Mesh (accurate)} for final numbers and STL export; use \emph{Voxel wireframe (preview)} for fast, memory light inspection on large datasets or when you only need a quick OBJ wireframe.

\noindent \textbf{Basic Steps:}
\begin{enumerate}
    \item Import the 3D stack in Fiji and convert to a binary image stack (as shown in \textbf{Figure~\ref{fig:img1}})
    \item Navigate to Plugins \,$\rightarrow$\, \texttt{StrengthLawExtractor} and a window as shown in \textbf{Figure~\ref{fig:img2}} will pop up 
    
    \item Set \emph{Pores appear as} (black/white), tolerance $\varepsilon$ (start with $0.00$ for binary), voxel sizes $(s_z,s_y,s_x)$ and units (use pixel, px, if actual distance not known), optionally enable \emph{Pad solid boundary}, choose \emph{Computation mode}, then click \emph{Compute Features}. 
    
    \item Use \emph{Export Results + STL} (or OBJ) to save outputs.
\end{enumerate}
\end{flushleft}

\section{Method and pipeline description}
\begin{flushleft}
This method converts a stack of 2D slices into a metrically correct 3D representation of the pore phase by rescaling voxel dimensions according to the physical spacing in each axis. In this way, the reconstruction preserves true geometric proportions rather than assuming isotropic voxels. From this 3D volume, it extracts the solid interface of the pores, and reports porosity, surface area, mean curvature and the Euler characteristic. All geometric quantities are evaluated in world coordinates using the per axis voxel spacing $\mathbf{s}=(s_z,s_y,s_x)$ so the areas and curvatures reflect acquisition anisotropy.

\paragraph{Data, units, and physical coordinates.}
The volumetric image is an intensity field $I:\Omega\subset\mathbb{Z}^3\to\mathbb{R}$, indexed by $\mathbf{i}=(z,y,x)$. The physical coordinates are $\mathbf{x}=\mathrm{diag}(\mathbf{s})\,\mathbf{i}=(s_z z,\ s_y y,\ s_x x)$. That is, voxel indices $\mathbf(z,y,x)$ are converted into real physical positions by multiplying with the voxel spacing $\mathbf{s}=(s_z,\ s_y,\ s_x)$. In this way, voxel counts become distances in calibrated units (e.g., $\mu$m or mm). Whenever referring to lengths, areas, or curvature integrals, they are taken after this scaling, so that all reported geometric quantities correspond to physical measurements in physical space rather than raw voxel coordinates.

\paragraph{Binary mapping with explicit phase convention.}
The convention $0=\text{pore}$ and $1=\text{solid}$ is fixed without data driven thresholds. Let $I_{\min}$ and $I_{\max}$ be the global extrema and let $\varepsilon\in[0,0.1]$ be a small fractional tolerance to absorb near extreme noise. If pores are darker than the solid (\textbf{Figure~\ref{fig:img3}}),
\[
\tau_{\mathrm{low}}=I_{\min}+\varepsilon\,(I_{\max}-I_{\min}),\qquad
B(\mathbf{i})=\begin{cases}
0,& I(\mathbf{i})\le\tau_{\mathrm{low}},\\
1,& I(\mathbf{i})>\tau_{\mathrm{low}}.
\end{cases}
\]

% where \(\tau_{\mathrm{low}}\) is X and \textit{B}(i) is X. 
where, $\tau_{\mathrm{low}}$ is the intensity threshold separating pores from solid, and $B(\mathbf{i})$ is the resulting binary mask at voxel $\mathbf{i}$, with $0=\text{pore}$ and $1=\text{solid}$.

If pores are brighter (as shown in \textbf{Figure~\ref{fig:img4}}), the rule is mirrored to preserve $0=\text{pore}$:
\[
\tau_{\mathrm{high}}=I_{\max}-\varepsilon\,(I_{\max}-I_{\min}),\qquad
B(\mathbf{i})=\begin{cases}
0,& I(\mathbf{i})\ge\tau_{\mathrm{high}},\\
1,& I(\mathbf{i})<\tau_{\mathrm{high}}.
\end{cases}
\]
This produces a strict $\{0,1\}$ field aligned with the stated polarity.

\begin{figure}[htbp]
  \centering
  \begin{subfigure}{0.49\linewidth}
    \centering
    \includegraphics[width=\linewidth]{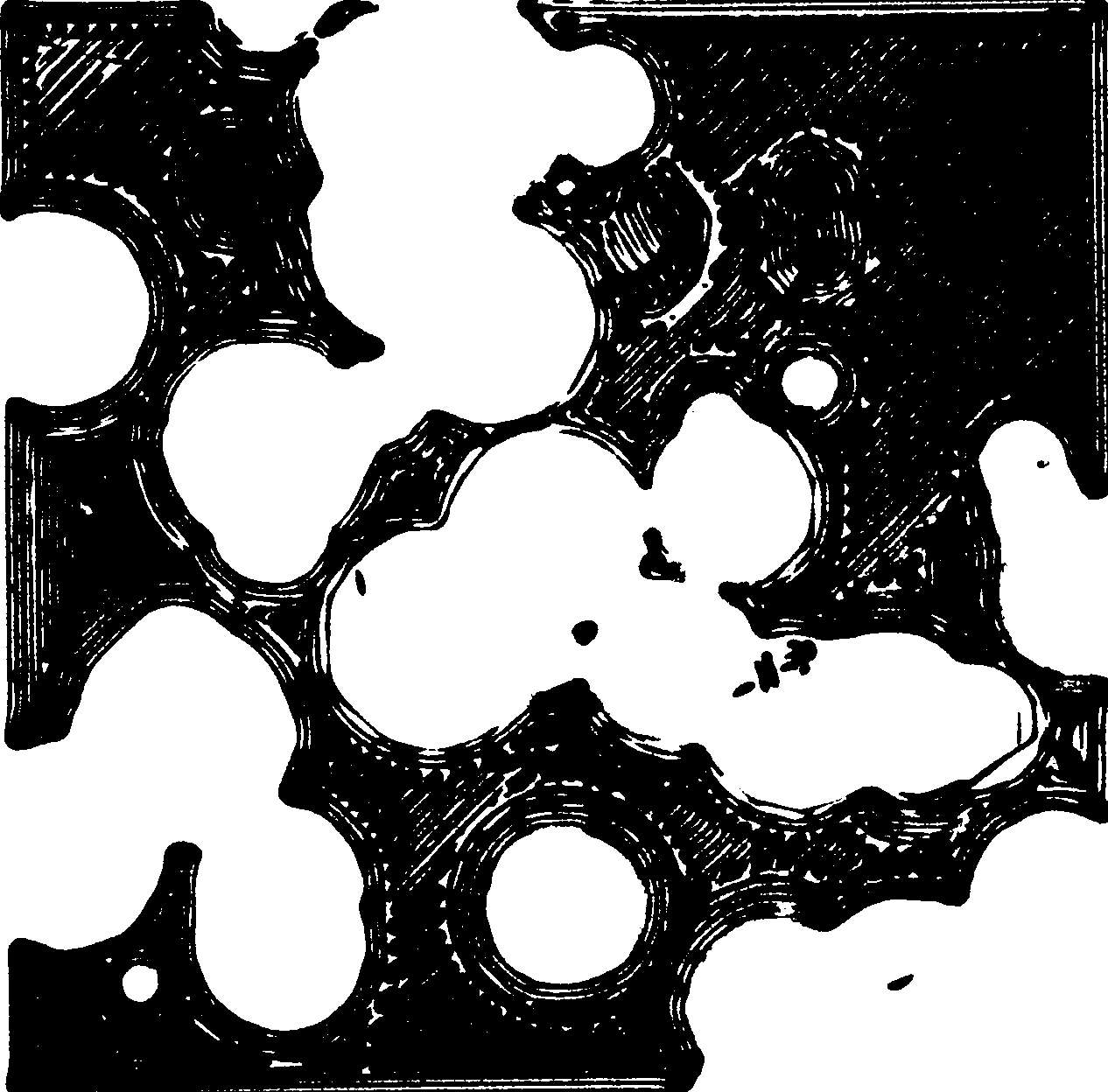}
    \caption{ }
    \label{fig:img3}
  \end{subfigure}\hfill
  \begin{subfigure}{0.49\linewidth}
    \centering
    \includegraphics[width=\linewidth]{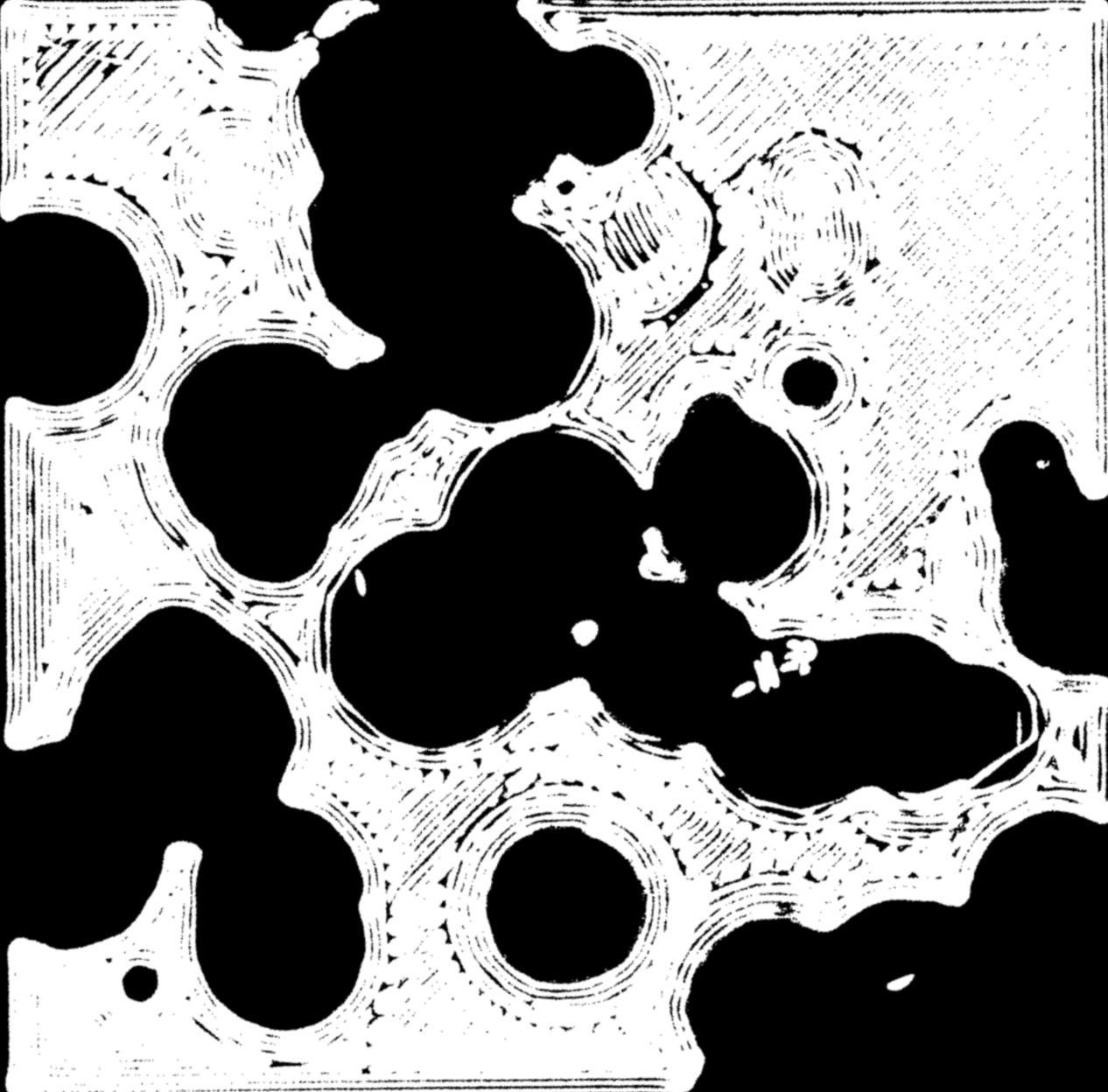}
    \caption{ }
    \label{fig:img4}
  \end{subfigure}
  \caption{(a) Example for pores are brighter; (b) Example for pores are darker than the solid.}
  \label{fig:imgs}
\end{figure}

\paragraph{Optional 3D morphological stabilization.}
To reduce noise and single voxel bridges without changing topology by default, conservative pore opening and solid closing with spherical structuring elements of radii $(r_o,r_c)$ voxels are provided. The defaults $r_o=r_c=0$ leave the binary field unchanged.

\paragraph{Boundary handling.}
Interfaces that touch the domain boundary bias curvature and area. An optional one voxel \emph{solid} padding $B_{\mathrm{pad}}$ closes open surfaces at the box and stabilizes dihedral angle estimates; disabling padding is used to preserve the original domain. For periodic media, the binary field can be tiled prior to interface extraction to enforce continuity across opposite faces and remove artificial boundary edges.

\paragraph{Two geometry paths.}
The pipeline supports two consistent paths:

\emph{(i) Mesh path.} The pore foreground $P=\mathbb{1}[B=0]$ is triangulated with marching cubes at isolevel $0.5$, yielding a watertight mesh $(V,F)$. In the reference implementation, vertices are produced in index units and scaled by $\mathbf{s}$ when evaluating geometry and when exporting; this yields the same metrics as supplying $\mathbf{s}$ to the extractor. For a face $f=\{i,j,k\}$ with world space vertices $v_i,v_j,v_k$, the area contribution is
$
A_f=\tfrac12\|(v_j-v_i)\times(v_k-v_i)\|, \quad A=\sum_{f\in F}A_f.
$
The mean curvature is evaluated from exterior dihedral angles on edges
\[
M \approx \tfrac12 \sum_{e\in E} \ell_e\,(\pi-\theta_e),
\]
with $\ell_e$ the world space edge length and $\theta_e$ the angle between unit face normals of the two incident triangles. The area normalised mean curvature is $\bar{H}=M/A$. The mesh Euler characteristic is $\chi=V-E+F$ after consolidating duplicate edges. Padding reduces nonmanifold edges on open interfaces and improves the stability of $M$.

\emph{(ii) Voxel (preview) path.} For fast, memory light previews, geometry is computed directly from the 3D cubical complex of the pore voxels, without meshing. Let $c$ denote counts of pore voxels and their shared elements: $c.n_3$ (voxels), $c.n_{2x},c.n_{2y},c.n_{2z}$ (adjacent voxel pairs along $x,y,z$), $c.n_{1xy},c.n_{1yz},c.n_{1zx}$ (filled $2{\times}2$ squares in the coordinate planes), and $c.n_0$ (filled $2{\times}2{\times}2$ cubes). Closed form expressions then yield
\[
\begin{aligned}
A &= (2c.n_3{-}2c.n_{2x})\,s_y s_z
   + (2c.n_3{-}2c.n_{2y})\,s_x s_z
   + (2c.n_3{-}2c.n_{2z})\,s_x s_y,\\
M &= \tfrac{\pi}{2}\,\bigl(c.n_{1yz}\,s_x + c.n_{1zx}\,s_y + c.n_{1xy}\,s_z\bigr),\\
\chi &= c.n_3 - (c.n_{1xy}+c.n_{1yz}+c.n_{1zx}) - c.n_0,
\end{aligned}
\]
with the isotropic special cases obtained by setting $s_x=s_y=s_z=s$. These formulas agree with mesh based values on rectilinear voxel surfaces and avoid holding a triangle mesh in memory.

\paragraph{Descriptors and conventions.}
Porosity is the 3D pore volume fraction $\phi=\#\{\mathbf{i}:B(\mathbf{i})=0\}/|\Omega|$ and does not depend on $\mathbf{s}$. Unless otherwise stated, Euler characteristic for the \emph{voxel} foreground uses 26/6 neighbor connectivity (object voxels 26 connected, background voxels 6 connected, standard for 3D data on a cubical grid) \parencite{Legland_2011}, whereas the \emph{mesh} $\chi$ follows $V{-}E{+}F$. All reported quantities inherit the physical units implied by $\mathbf{s}$.

\paragraph{Numerical safeguards.}
All solid or all pore inputs yield $A=0$ and undefined curvature; the implementation returns $A=0$, $M=\mathrm{NaN}$, and $\bar{H}=\mathrm{NaN}$. Small $\varepsilon$ values in the range $0.00$–$0.02$ typically absorb near extreme noise without shifting the phase boundary; larger values should be used with caution. However, for a binarized stack input, $\varepsilon$ can be set to $0$ as the binarized input will be only $0$ or $1$ for dark or light. Gentle morphology and optional padding improve geometric stability on noisy or boundary touching interfaces while preserving the fixed phase convention.

\paragraph{Outputs and quality control.}
For external validation, the strict $\{0,1\}$ stack can be exported as a TIFF stack file for slice wise inspection. If connectivity sanity checks are needed, a 3D skeleton of the pore mask (endpoints and junctions) can be generated to summarize network structure. The interface geometry can be exported either as STL (mesh path) or as OBJ wireframes: from mesh edges or directly from exposed voxel faces in the preview path. All exports are scaled by $(s_z,s_y,s_x)$ so that coordinates, areas, and curvatures are in physical units.

\paragraph{Reproducibility.}
Given the image stack, voxel spacing $\mathbf{s}$, tolerance $\varepsilon$, optional morphology $(r_o,r_c)$, and boundary choice (none, padding, or periodic tiling), all steps are deterministic. The fixed phase convention $0=\text{pore}$ removes ambiguity across datasets.

\end{flushleft}

\section{Implementation}
\begin{flushleft}
    
The \texttt{StrengthLawExtractor} Fiji plugin exposes the pipeline through a compact dialog. Users open a 3D stack, specify the pore polarity (darker or brighter than solid), set the fractional tolerance $\varepsilon$, enter voxel sizes $(s_z,s_y,s_x)$ with a units label, and choose whether to pad the solid boundary. A computation mode selector offers \emph{Mesh (accurate)} or \emph{Voxel wireframe (preview)}. A progress dialog provides percentage feedback during conversion, meshing, geometry evaluation, and export.

\begin{figure}[htbp]
  \centering
  \includegraphics[width=0.95\textwidth]{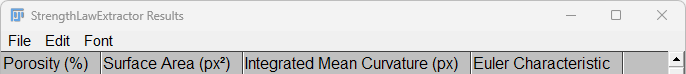}
  \caption{Result report window.}
  \label{fig:plugin_gui}
\end{figure}

In mesh mode, the binary foreground is triangulated at isolevel $0.5$ and the descriptors $(\phi,A,M,\chi)$ are evaluated on the world scaled mesh, with $\bar{H}=M/A$ available for convenience. In preview mode, the mesh is omitted and the same descriptors are computed from voxel counts using the closed form expressions above. Results are displayed in the \texttt{Results Table} of Fiji (as shown in \textbf{Figure~\ref{fig:plugin_gui}}) and can be written to \texttt{.csv}. Geometry export supports STL of the mesh (as shown in \textbf{Figure~\ref{fig:plugin_gui2}}), OBJ wireframe derived from mesh edges, and OBJ wireframe traced directly from voxel boundary faces for preview mode. All writers apply the physical scaling $\mathbf{s}$ so saved coordinates match the reported units.

\begin{figure}[htbp]
  \centering
  \begin{subfigure}{0.479\linewidth}
    \centering
    \includegraphics[width=\linewidth]{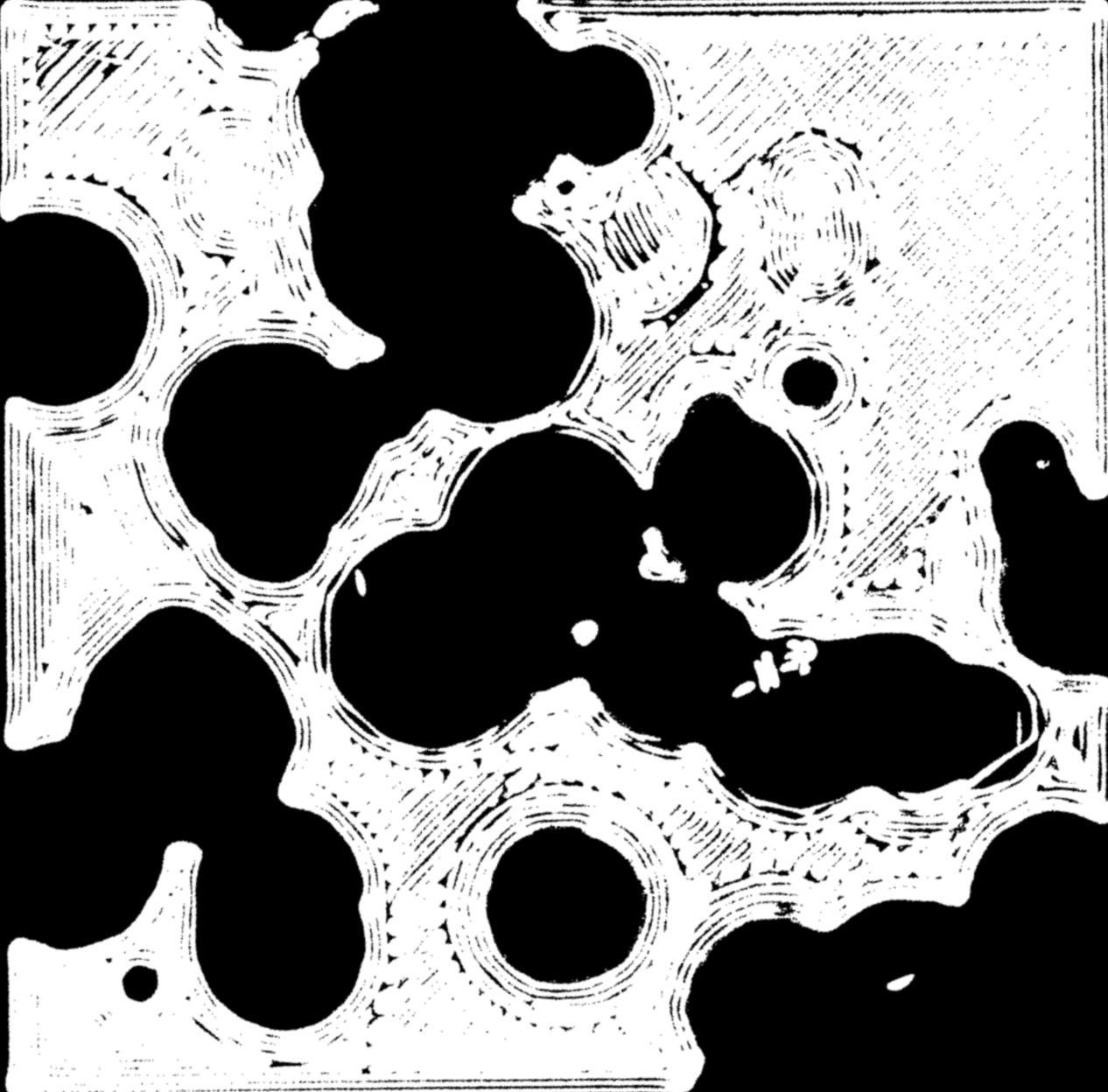}
    \caption{ }
    \label{fig:input-slice}
  \end{subfigure}\hfill
  \begin{subfigure}{0.515\linewidth}
    \centering
    \includegraphics[width=\linewidth]{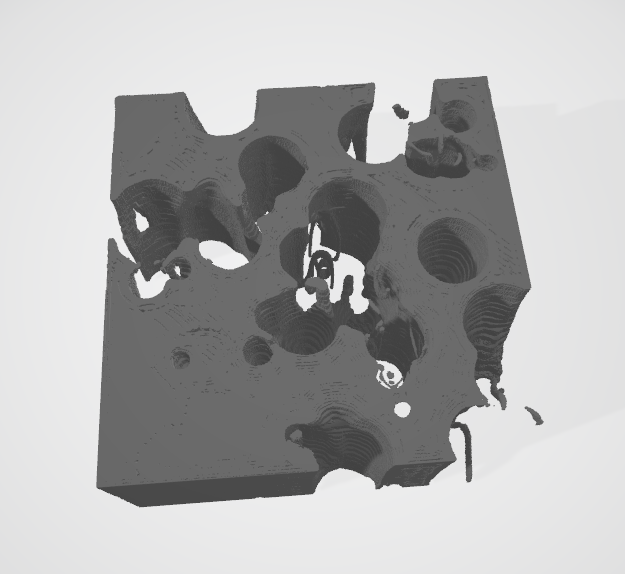}
    \caption{ }
    \label{fig:output-mesh}
  \end{subfigure}
  \caption{(a) A single micro-CT scan slice (cross-section) used as input; (b) the corresponding output mesh.}
  \label{fig:plugin_gui2}
\end{figure}

\end{flushleft}

% \section{Results and Demonstration}
% \subsection{Validation and Comparison}

\section{Results}
\begin{flushleft}
\begin{itemize}[leftmargin=*, label={}, topsep=0pt, itemsep=0pt]
\item{\textbf{Methods under comparison.} }
Four descriptors were evaluated: porosity, surface area, mean curvature, and Euler characteristic. Measurements were produced using four method variants: \texttt{StrengthLawExtractor} in mesh mode, \texttt{StrengthLawExtractor} in voxel mode, BoneJ \parencite{Doube2010BoneJ}, and MorphoLibJ \parencite{10.1093/bioinformatics/btw413}. The mesh mode computes geometry on a watertight triangular interface and produces exportable meshes. The voxel mode uses closed form estimators on the cubical complex and is suited to memory efficient previews on large volumes. BoneJ and MorphoLibJ provide porosity, surface related measures, and connectivity metrics; mean curvature is not available in either toolkit.

\item{\textbf{Evaluation protocol.}}
All tools operated on identical binary volumes derived from the same micro-CT stacks, with the same voxel spacing supplied to ensure unit consistent values. Unless otherwise stated, padding was disabled, morphology was not applied, and the fractional tolerance for threshold mirroring was kept small to absorb near extreme noise. Surface area is reported in squared physical units from the voxel spacing (for example mm$^2$). Mean curvature is reported in the corresponding length unit (for example mm). Porosity and Euler characteristic are unitless. All methods were run on a computer with 128 GB of RAM and the same version and RAM settings of Fiji \parencite{Schindelin2012Fiji}.

\item{\textbf{Datasets.}}
% There are two sets of dataset prepared. One is a 
A binary three dimensional stack containing a single solid sphere was created in Fiji ImageJ using 3D ImageJ Suite \(\to\) Shapes3D \(\to\) Sphere. The stack size was \(512\times512\times512\) voxels with calibration \(1\times1\times1\) in unit px (as shown in \textbf{Figure~\ref{fig:data}}). The sphere center was \((256,256,256)\) voxels and the radius was \(64\) voxels. Foreground voxels were set to \(255\) and background voxels to \(0\), and the sphere was fully contained inside the field of view. In the \textit{Table 1}, the datasets are listed as "Sphere".

\begin{figure}[htbp]
  \centering
  \includegraphics[width=0.6\textwidth]{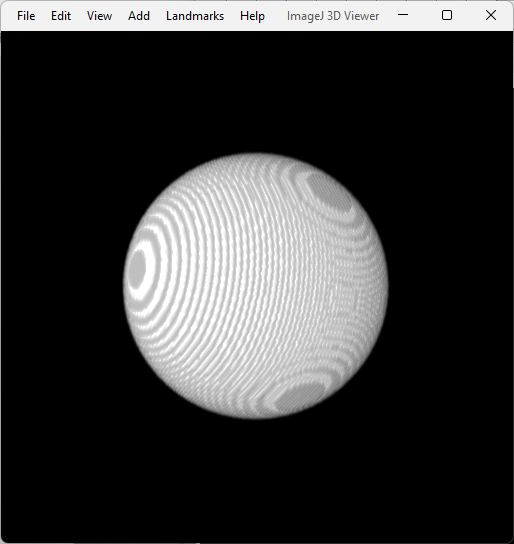}
  \caption{A 3D visualization of the synthetic sphere by the 3D viewer in Fiji.}
  \label{fig:data}
\end{figure}

\end{itemize}
\end{flushleft}   

\subsection{Comparative Evaluation}
\begin{flushleft}
The four morphometers, porosity, surface area, mean curvature, and Euler characteristic, were compared across the \texttt{StrengthLawExtractor} plugin (mesh and voxel paths), BoneJ, and MorphoLibJ on the two datasets described above. All tools were applied to identical binary volumes with the same voxel spacing to ensure direct comparability. Porosity is unitless. Surface area is reported in squared physical units determined by the voxel spacing (e.g., mm$^2$). Mean curvature is reported in the corresponding length unit (e.g., mm). BoneJ do not provide mean curvature and and MorphoLibJ do not provide the four features directly; those not applicable entries are marked “n/a.” The mesh path of \texttt{StrengthLawExtractor} plugin is designed to evaluate geometry on a watertight triangular interface and produce exportable meshes, whereas the voxel path uses closed form estimators that agree with the mesh on rectilinear voxel surfaces and reduce memory and runtime on large volumes. The STL output from StrengthLawExtractor for the sphere dataset is shown in \textbf{Figure~\ref{fig:data_out_stl}}.

\paragraph{How the four features were obtained}
MorphoLibJ $\rightarrow$ Analyze $\rightarrow$ Microstructure Analysis was run with Volume, Surface Area, Mean Breadth, and Euler Number selected, Crofton with 13 directions for surface area and mean breadth, and standard connectivity. The result table headers are exactly
\[
\texttt{VolumeDensity}\,,\ \texttt{SurfaceAreaDensity}\,,\ \texttt{MeanBreadthDensity}\,,\ \texttt{EulerNumberDensity}.
\]
These quantities are densities, therefore conversion to totals uses the image volume $V=N_x N_y N_z\, v_x v_y v_z$:
\[
\mathrm{Porosity}=100\,(1-\mathrm{VolumeDensity}),\qquad
S=S_v\,V,\qquad
\int_{\partial K} H\,\mathrm{d}A=2\pi\,B_v\,V,\qquad
\chi=E_v\,V,
\]
where $S_v=\texttt{SurfaceAreaDensity}$, $B_v=\texttt{MeanBreadthDensity}$, and $E_v=\texttt{EulerNumberDensity}$.

\paragraph{Sphere dataset}
Voxel size $1$, $V=512^3$.  
Output $\texttt{0.008\;\;3.900E{-}4\;\;9.664E{-}7\;\;7.494E{-}9}$.  
Porosity $99.200\,\%$, $S=52{,}344.914\ \mathrm{px}^2$, $\displaystyle \int H\,\mathrm{d}A=814.979\ \mathrm{px}$, $\chi\approx 1$.

% \paragraph{Cube dataset}
% Calibrated stack with volume $V$ from stack dimensions and voxel sizes.  
% Output $\texttt{0.504\;\;8.112\;\;-8.415\;\;610.260}$.  
% $\mathrm{Porosity}=49.6\,\%$, $S=8.112\,V$, $\displaystyle \int H\,\mathrm{d}A=2\pi(-8.415)\,V$, $\chi=610.260\,V$.

\begin{table}[htbp]
  \centering
  \caption{Four morphometers comparison across datasets. StrengthLawExtractor is shortened as SLE to save space. Porosity is unitless; surface area in [\textit{unit}$^2$]; mean curvature in [\textit{unit}]. “n/a” indicates a feature not provided by that toolkit.}
  \label{tab:multi-dataset-four-feature}
  \begingroup\small
  \setlength{\tabcolsep}{12pt}
  \begin{tabular}{l l r r r r}
    \toprule
        Dataset & Feature & SLE (Mesh) & SLE (Voxel) & BoneJ & MorphoLibJ \\ 
    \midrule
    \multirow{4}{*}{\textit{Sphere}} 
      
      & Porosity [ \textit{\%} ]                         & 99.17 & 99.17 & 99.17 & 99.20  \\
      & Surface area [\textit{unit}$^2$]  & 56605.31 & 56605.31 & 55911.78 & 52344.92   \\
      & Mean curvature [\textit{unit}]    & 1215.796 & 1215.796 & n/a & 814.98   \\
      & Euler characteristic              & 1  & 1 & 1 & 1   \\

    \bottomrule
  \end{tabular}
  \endgroup
\end{table}

Results indicate close agreement among methods for porosity and Euler characteristic when identical binaries and spacing are used. Surface area exhibits small, expected differences among implementations due to discretization and boundary handling; values from the mesh and voxel paths remain within tight deltas on these volumes. Mean curvature is available only from \texttt{StrengthLawExtractor}, computed either on a watertight mesh (mesh path) or via closed form voxel estimators (voxel path). Because all quantities are evaluated in world coordinates using the supplied voxel spacing, anisotropic acquisitions are handled consistently: areas scale with face areas and mean curvature with edge lengths in the chosen physical unit.

\begin{figure}[htbp]
  \centering
  \includegraphics[width=0.6\textwidth]{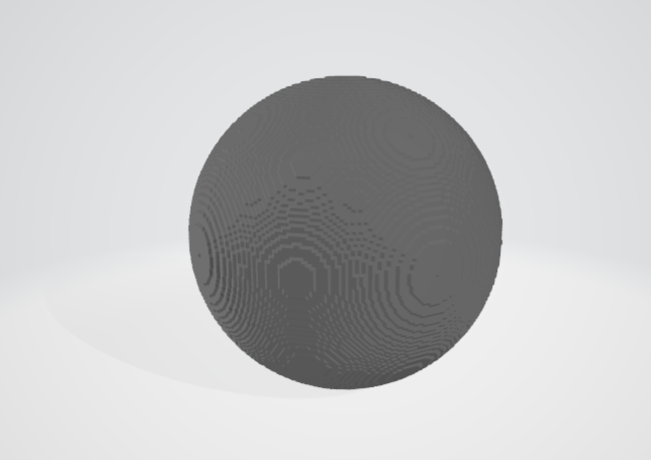}
  \caption{STL output produced by StrengthLawExtractor for the sphere dataset.}
  \label{fig:data_out_stl}
\end{figure}

\end{flushleft}

\section{Discussion and Limitations}
\begin{flushleft}
The development of non-destructive approaches for characterizing porous materials represents a critical step in bridging the gap between imaging and predictive modeling efficiently. The presented Fiji plugin addresses the need for standardized quantification of the four morphometric measures that govern strength laws, providing a direct pathway from micro-CT data to descriptors that can be incorporated into constitutive models or machine learning workflows. By automating feature extraction within a widely adopted open-source platform, the plugin reduces user bias, improves reproducibility, and lowers the barrier for integrating morphometric analysis into materials research.

Despite these advantages, several limitations remain. The accuracy of morphometric measures depends strongly on the resolution and quality of the X-ray micro-CT scans, and low resolution data may obscure fine scale structural details critical to mechanical behavior. The current implementation is optimized for typical laboratory datasets and may require adaptation for very large volumes or for data obtained from synchrotron facilities with higher dimensionality. Furthermore, the plugin focuses on geometric descriptors and does not directly capture physical processes such as mineral dissolution, crack propagation, or phase transformations, which may also contribute to material strength. Integration with constitutive modeling therefore still requires assumptions about the mapping from morphometry to mechanical response. 

When comparing the presented plugin with BoneJ and MorphoLibJ, differences in design choices are noticed. BoneJ applies a 26/6 connectivity rule, where objects are considered 26 connected and the background 6 connected. Surface area and mean curvature are calculated from marching cube meshes, and the Euler characteristic is consistently derived using this convention. MorphoLibJ, in contrast, uses connectivity only in its two dimensional Crofton estimator, which can be set to either 4 or 8 connectivity. It relies on voxel based Crofton or lookup table methods instead of meshing, which makes its estimates distinct from those produced by BoneJ. The plugin described here also adopts the 26/6 convention but offers two complementary modes: a voxel count mode, based on exact Minkowski formulas that provide robust Euler numbers and voxel faithful geometry, and a mesh mode using marching cubes with dihedral angles, which produces results closer to BoneJ. These differences do not represent shortcomings of any single approach, but rather reflect alternative strategies for balancing precision, scalability, and computational efficiency. Together, the differences highlight the importance of understanding methodological assumptions when interpreting morphometric results across different software tools.

Finally, while the plugin facilitates descriptor extraction, the predictive capability ultimately depends on the availability of robust experimental or simulated datasets for model calibration.

\end{flushleft}

\section{Conclusion and Future Work}
\begin{flushleft}
This study introduced a Fiji plugin, \texttt{StrengthLawExtractor}, designed to compute four essential morphometric descriptors from X-ray micro-CT scans of porous materials, enabling a non-destructive and reproducible route to link microstructure with macroscopic mechanical behavior. By providing automated feature extraction within an accessible framework, the plugin advances the practical application of morphometric strength laws and supports integration with emerging data driven approaches. The tool has broad applicability across geoscience, materials science, and biomechanics, offering a pathway toward more systematic material evaluation and design.

Future work will focus on extending the plugin to handle larger datasets and improve computational efficiency, allowing application to high resolution synchrotron and industrial scale imaging. Additional functionality may include integration with supervised and unsupervised machine learning models, enabling direct prediction of stress-strain curves from imaging data. Incorporation of physics informed descriptors, such as measures of crack networks or dissolution patterns, could further enhance the predictive accuracy. Finally, coupling the plugin with open databases of porous microstructures would promote reproducibility and accelerate the development of generalized models for porous media mechanics.
\end{flushleft}

\section{Availability and Requirements}
\begin{flushleft}

\textbf{Software.} \texttt{StrengthLawExtractor} is available as an open source Fiji plugin. Source code and prebuilt releases are provided at \parencite{qyt21_2025_zenodo_17394095} under the \textit{Apache-2.0} license.

\textbf{Platform.} Runs on Windows. A 64 bit Java Runtime Environment is required. (Not yet tested on macOS and Linux.)

\textbf{Dependencies.} Fiji distribution with ImageJ2 \parencite{Rueden2017} and ImgLib2 \parencite{10.1093/bioinformatics/bts543}. The plugin uses ImageJ Ops \parencite{Rueden2021} for meshing and geometry. 
% Tested with Fiji \textit{2025}, Java \textit{version}.

\textbf{Hardware.} Validation was performed on a workstation with 128 GB of RAM. Actual memory use depends on dataset size and chosen mode. Mesh mode provides exportable interface geometry and may require substantial RAM for large volumes. When available memory is limited, voxel mode offers an alternative that reduces memory use while providing the same four global descriptors.

\textbf{Installation.} Either enable the \textit{Update Site} in the Fiji updater and restart, or place the provided \texttt{StrengthLawExtractor.jar} file in the \texttt{Fiji.app/plugins/} folder located in your computers files and restart Fiji.
\end{flushleft}

\vspace{-5mm}
\begin{flushleft}

\end{flushleft}

\begin{flushleft}\textbf{Funding}\\
The Duke University Department of Civil and Environmental Engineering is greatly acknowledged for their support. 
\end{flushleft}

\begin{flushleft}\textbf{Acknowledgments}\\
The Duke University Department of Civil and Environmental Engineering is greatly acknowledged for their support. 
\end{flushleft}

\begin{flushleft}\textbf{Data and Code Availability}\\
See Section 8.

\end{flushleft}

% \bibliographystyle{unsrtnat}
%\bibliographystyle{abbrvnat}
% \bibliographystyle{plainnat}

% \bibliography{plugin-ref}
% \printbibliography[title={References}]
\printbibliography

\end{document}